# Circles: Inter-Model Comparison of Multi-Classification Problems with High Number of Classes

Nina Mir*  
San Francisco State University, USA.

Ragaad AlTarawneh†  
Intel Labs, USA

Shah Rukh Humayoun‡  
San Francisco State University, USA

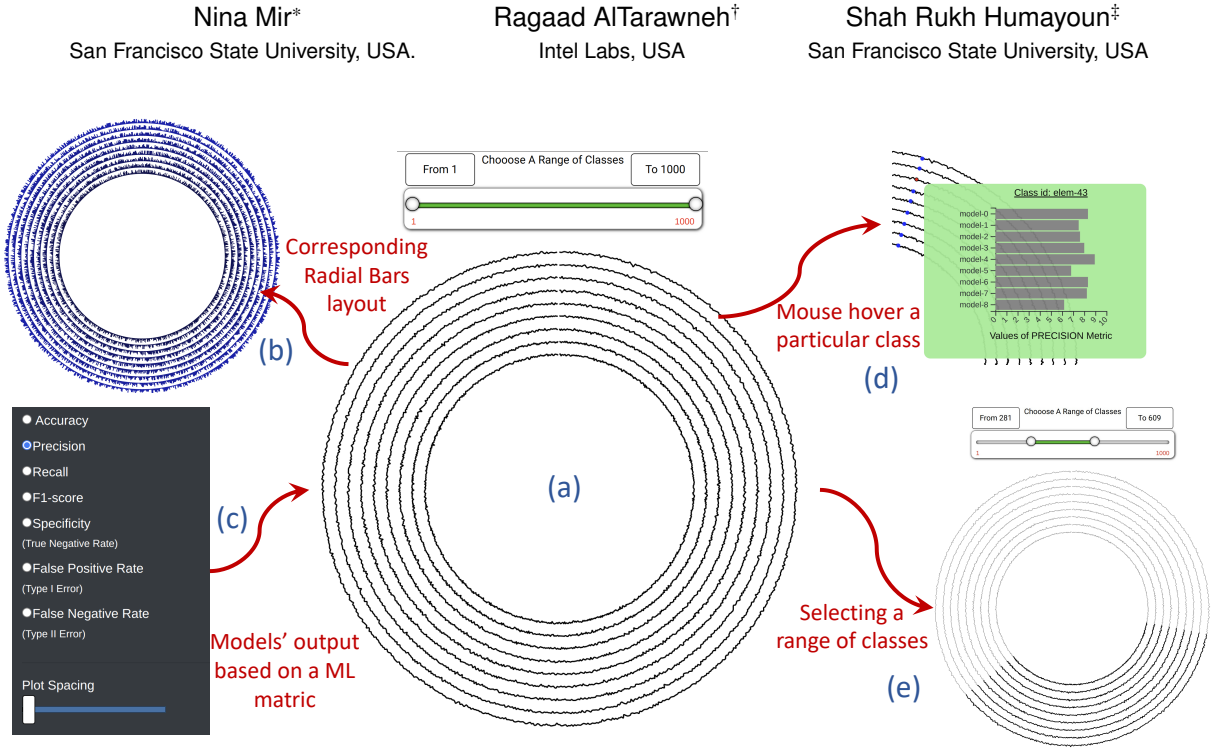

Figure 1: **(a)** Circles inter-model comparison view, where all classification models' outputs are displayed in a concentric radial line view. **(b)** Showing the models' outputs using a radial bar chart. **(c)** *Metric Selection Panel* provides the option to select a ML metric to be used for model comparison view in (a). **(d)** Mouse hover a particular class highlights the same class in all models and a tool-tip appears to show the value of used ML metric using horizontal bar chart. **(e)** Range slide bar provides the option to highlight only the classes within the selected range.

## Abstract

The recent advancements in machine learning have motivated researchers to generate classification models dealing with hundreds of classes such as in the case of image datasets. However, visualization of classification models with high number of classes and inter-model comparison in such classification problems are two areas that have not received much attention in the literature, despite the ever-increasing use of classification models to address problems with very large class categories. In this paper, we present our interactive visual analytics tool, called **Circles**, that allows a visual inter-model comparison of numerous classification models with 1K classes in one view. To mitigate the tricky issue of visual clutter, we chose concentric a radial line layout for our inter-model comparison task. Our prototype shows the results of 9 models with 1K classes in a single view; however, up to 20 models' results can be displayed in this way.

*e-mail: ninamirf@gmail.com  
†e-mail: ragaad.altarawneh@intel.com  
‡e-mail: humayoun@sfsu.edu

## 1 Introduction

With the rise of machine learning (ML) as a service model, selecting a classifier, training it and running test data without having expert-level knowledge about the inner workings of these models have become a reality. In last few years, we have seen development of various public and commercial platforms using this approach such as Google Cloud AutoML [2], Microsoft Azure Machine Learning, Amazon Machine Learning Services, etc. These recent advancements in ML allow data analysts to use numerous models to tackle their problems. However, they face the issue of selecting the best model suitable to their domain and problem at hand [3].

Many visual analytics (VA) tools have been proposed to do comparison of multi-class classification models; however, mostly they deal with only a few classes, e.g., [11] demonstrated a scenario with 10 classes while [9] handles no more than 40 classes (film genres). In the case of datasets with hundreds of classes, such as ImageNet ILSVRC [12] dealing with 1K classes of images, it becomes a challenging issue to compare different models, even for seasoned ML analysts. Although some solutions have been proposed to visualize such models, they focus on only showing one model's results, e.g.: Alsallakh et al. [1] showed the results of one classifier model via a matrix visualization, Uwaseki et al. [14] used a confusion matrix layout to display results of a single image classifier model, and Ono et al. [9] used a concentric radial view to visualize the results of one multi-task classifier. Visualizing models with hundreds of classes requires intuitive solutions to enable the users to explore and compare models so to select the best model for their problem at hand.

Targeting this concern, we present an interactive visual analytics tool, called **Circles**, that allows users to compare results of multiple multi-class classification models, targeting 1K classes in ILSVRC dataset, using an interactive concentric radial view. The Circles tool allows users to explore and compare between different models using the most common used ML metrics in classification problems.

## 2 THE DATASET

ImageNet [4] dataset consists of a collection of over 15 million high resolution labelled images, belong to an estimated 22k different categories. We use ImageNet Large-Scaled Visual Recognition Challenge (ILSVRC) dataset [12] that is a subset of ImageNet dataset with images belong to 1k different categories. Overall, ILSVRC contains an estimated 1.2 million training images, 50k validation images and 150k testing images. Each model output is a 2-dimensional vector space where the prediction distribution is distributed across 1K classes and the output is saved in JSON format for portability.

## 3 THE CIRCLES TOOL

Our developed tool, called **Circles**, visualizes results of multiple multi-class classification models targeting 1K classes in ILSVRC [12] dataset. The web-based client side was developed using HTML, CSS, JavaScript, D3.js library, and Vue framework. The server side was developed using Node.js run-time environment to manage and process the imported data.

One of the main challenges in our target multi-class classification problem is displaying the results of 1K classes. In the case of few classes or a single model, a vertical layout (e.g., [11]) or a matrix layout (e.g., [5, 6, 10]) could work. However, we chose a radial (circular) layout approach (see Fig. 1) to inter-model comparison, as they produce compact visualizations and use space efficiently [7]. This is because they support a larger data domain on a square area compared to rectangular or square layouts [7, 8]. Also, they encourage the eye movement to proceed along the curved lines rather than a zig-zag fashion in a square or rectangular figure, which helps users to better understand and explore the underlying data [8].

In Circles, users import the target classification models' JSON files to compare them visually. Circles computes and visualizes these models using seven most common used ML metrics [13] of interest in classification problems, i.e., *accuracy*, *precision*, *recall*, *F1-score*, *specificity*, *false positive rate*, and *false negative rate*. These metrics are computed, using the formulas provided in [13], at class-level. Once the user selects a particular metric, Circles renders the underlying models in a concentric radial view using radial (circular) lines. In Fig. 1(a), each radial line represents one of the input models, where each model radial line is plotted using the prediction distribution of validation images in ILSVRC dataset using one of the selected ML metrics against 1K classes.

The lower and upper bound for each of the ML metrics are, respectively, 0 and 1. This bound is projected into a circular buffer area of 10 pixels (0.1 = 1 px). In other words, each model's radial curve is plotted within this circular buffer area. However, we have added more than 10 pixels of distance between the concentric curves to increase visibility and to avoid visual clutter. Further spacing can be achieved by using the Plot Spacing slider (Fig. 1(c)).

We use radial line path generator in D3.js library, which works with radial coordinates, to create these path elements. In order to accommodate 1K classes, one revolution (360 degrees) is divided into 1000 equal parts, resulting in 1000 non-overlapping points around a circle. All that is left is deciding the separation distance (pixels) between the resulting concentric circular paths to avoid inter-model overlapping. Further, we use D3.js library interpolator to produce a cubic Catmull-Rom spline, which not only generates smooth curves but also ensures that the resulting path goes through all the controlled points. Circles also provides the option to show the view in radial bars based on demand (Fig. 1(b)).

Circles provides several interaction and filtering options for better inter-model comparison, e.g., users can change the space between radial lines using a slider (Fig. 1(c)). Further, a range slider is provided to select a portion of classes, which results in highlighting only the classes in range (Fig. 1(e)). Mouse hover a particular class in any model highlights it in red color and also highlights the same class in all other models in blue color (Fig. 1(d)). Furthermore, a tooltip is appeared to show the value of used ML metric of this class in all models through horizontal bar chart (Fig. 1(d)).

## 4 FUTURE WORK

The presented Circles tool enables the visual comparison of numerous classification models with high number of classes (e.g., 9 models with 1K classes are presented in this paper). In the future, we intend to execute a detailed user study targeting the inter-model comparison view. We also plan to conduct detailed user studies to gain more insight about the needs of ML analysts who work with classification models. Finally, we would like to provide additional facilities in Circles's inter-model comparison view, e.g., additional class filtration features, color-coding to add an additional dimension to our visuals, etc.